\documentclass[aps,prl,twocolumn,preprintnumbers,superscriptaddress,dblfloatfix,nofootinbib]{revtex4-1}
\usepackage[utf8]{inputenc}
\usepackage{lmodern}
\usepackage{amsmath,amssymb,mhchem}
\usepackage{graphicx}
\usepackage{url}
\usepackage{color}
\usepackage{enumitem}
\usepackage{subfigure}
\usepackage[dvipsnames]{xcolor}
\usepackage[colorlinks=true,breaklinks=true]{hyperref}
\hypersetup{allcolors=[rgb]{0.0 0.0 0.6},linkcolor=[rgb]{0.75 0.05 0.05}}
\usepackage{bm}
\usepackage{epsfig}
\usepackage{amsmath}
\usepackage{amssymb}
\usepackage{slashed}
\usepackage{color}
\usepackage{accents}
\usepackage{url}
\usepackage[dvipsnames]{xcolor}

\newcommand{\exclude}[1]{}

\begin{document}

\preprint{IPMU20-0076}
\preprint{RIKEN-iTHEMS-Report-20} 

\title{Constraining Primordial Black Holes with Dwarf Galaxy Heating}

\author{Philip Lu} \email{philiplu11@gmail.com}
\affiliation{Department of Physics and Astronomy, University of California, Los Angeles \\ Los Angeles, California, 90095-1547, USA} 

\author{Volodymyr Takhistov} \email{volodymyr.takhistov@ipmu.jp}
\affiliation{Department of Physics and Astronomy, University of California, Los Angeles \\ Los Angeles, California, 90095-1547, USA}
\affiliation{Kavli Institute for the Physics and Mathematics of the Universe (WPI), UTIAS \\The University of Tokyo, Kashiwa, Chiba 277-8583, Japan}

\author{Graciela B. Gelmini} \email{gelmini@physics.ucla.edu}
\affiliation{Department of Physics and Astronomy, University of California, Los Angeles \\ Los Angeles, California, 90095-1547, USA} 

\author{Kohei Hayashi} \email{k.hayasi@astr.tohoku.ac.jp}
\affiliation{Institute for Cosmic Ray Research, The University of Tokyo, Kashiwa 277-8582, Japan}
\affiliation{Astronomical Institute, Tohoku University, Sendai 980-8582, Japan}

\author{Yoshiyuki Inoue} \email{yinoue@astro-osaka.jp}
\affiliation{Department of Earth and Space Science, Graduate School of Science, Osaka University, Toyonaka, Osaka 560-0043, Japan}
\affiliation{Interdisciplinary Theoretical \& Mathematical Science Program (iTHEMS), RIKEN, 2-1 Hirosawa, Saitama 351-0198, Japan}
\affiliation{Kavli Institute for the Physics and Mathematics of the Universe (WPI), UTIAS \\The University of Tokyo, Kashiwa, Chiba 277-8583, Japan}

\author{Alexander Kusenko} \email{kusenko@ucla.edu}
\affiliation{Department of Physics and Astronomy, University of California, Los Angeles \\ Los Angeles, California, 90095-1547, USA}
\affiliation{Kavli Institute for the Physics and Mathematics of the Universe (WPI), UTIAS \\The University of Tokyo, Kashiwa, Chiba 277-8583, Japan}

\begin{abstract}
Black holes formed in the early universe, prior to the formation of stars, can exist as dark matter and also contribute to the black hole merger events observed in gravitational waves. We set a new  limit on the abundance of primordial black holes (PBHs) by considering interactions of PBHs with the interstellar medium, which result in the heating of gas. We examine generic heating mechanisms, including  emission from the accretion disk, dynamical friction, and disk outflows.  
Using the data from the Leo T dwarf galaxy, we set a new cosmology-independent limit on the abundance of PBHs in the mass range $\mathcal{O}(1) M_{\odot}-10^7 M_{\odot}$, relevant for the recently detected gravitational wave signals from intermediate-mass BHs.  
\end{abstract}

\date{\today}

\maketitle

\section{Introduction} \label{sec:intro}

Primordial black holes (PBHs) can form in the early Universe through a variety of mechanisms and can account for all or part of the dark matter (DM) (e.g.~\cite{Zeldovich:1967,Hawking:1971ei,Carr:1974nx,GarciaBellido:1996qt,Khlopov:2008qy,Frampton:2010sw,Kawasaki:2016pql,Carr:2016drx,Inomata:2016rbd,Pi:2017gih,Inomata:2017okj,Garcia-Bellido:2017aan,Georg:2017mqk,Inomata:2017vxo,Kocsis:2017yty,Ando:2017veq,Cotner:2016cvr,Cotner:2017tir,Cotner:2018vug,Sasaki:2018dmp,Carr:2018rid,Banik:2018tyb,1939PCPS...35..405H,Cotner:2019ykd,Kusenko:2020pcg,Flores:2020drq}). PBHs surviving until present can span many orders of magnitude in mass, from $10^{15}$~g to well over $10^{10}~M_{\odot}$, and they can account for the entirety of the DM in the mass window $\sim 10^{-16} - 10^{-10} M_{\odot}$, where there are no observational constraints~\cite{Katz:2018zrn,Smyth:2019whb,Montero-Camacho:2019jte}.
PBHs with sublunar masses can play a role in the  synthesis of heavy elements, production of positrons, as well as other astrophysical phenomena~\citep{Fuller:2017uyd,Takhistov:2017nmt,Takhistov:2017bpt,Takhistov:2020vxs}. PBHs with larger masses can account for some of the gravitational wave events detected by LIGO~(e.g. \cite{Abbott:2016blz,Abbott:2016nmj,Abbott:2017vtc}) as well as seeds of supermassive black holes~\cite{Bean:2002kx,Kawasaki:2012kn,Clesse:2015wea}.
The mass window of $10-10^3 M_\odot$ is particularly intriguing in connection with signals observed by LIGO~\cite{Nakamura:1997sm,Clesse:2015wea,Bird:2016dcv,Raidal:2017mfl,Eroshenko:2016hmn,Sasaki:2016jop,Clesse:2016ajp}. Very recently, LIGO announced first detection of an intermediate-mass $150 M_\odot$ BH \citep{Abbott:2020tfl}.
While a variety of constraints exist for this PBH mass range~(see Ref.~\citep{Carr:2020gox} for review), they often rely on multiple assumptions and are subject to significant uncertainties. 

In this work, we set new cosmology-independent constraints on PBH abundance based on the lack of gas  heating from PBH interactions with the  interstellar medium (ISM).  We consider several generic heating mechanisms, including dynamical friction, accretion disk emission as well as mass outflows/winds from accretion disk. We then apply our analysis to dwarf DM-rich galaxies, focusing on Leo T. Leo T is a transitional object between a dwarf irregular galaxy and a dwarf spheroidal galaxy that has been well studied and modeled theoretically. It has the  desired properties, such as a low baryon velocity dispersion,  making it a sensitive probe of PBH heating. While constraints on BHs interacting with surrounding stars have been extensively discussed~\citep{Carr:1999dyn,Lacey:1985,Totani:2010}, gas heating has not been considered in detail. Other constraints focused on the X-ray emission, but not the heating of the surrounding  gas~\citep{Gaggero:2016dpq,Inoue:2017csr}. ISM heating has been used to constrain particle DM candidates~\citep{Bhoonah:2018wmw,Farrar:2019qrv,Wadekar:2019xnf}, which have different heating mechanisms with a different velocity dependence compared to PBHs.  

Additionally, emission from PBH accretion can result in other observables and affect cosmological history, including modifications to reionization and recombination. As discussed in \cite{Ricotti:2007au}, ionization from X-rays due to PBH accretion will increase the amount of molecular hydrogen and the resultant cooling will enhance early star formation. PBH emission will also result in spectral distortions of the cosmic microwave background (CMB).
Furthermore, heating and ionization of intergalactic medium arising from PBH emission will modify 21 cm  power spectrum and will be probed by upcoming experiments \cite{Mena:2019nhm}. While we do not discuss these effects in detail within our work, they provide complementary probes of PBHs. 

\section{Black Holes in Interstellar Medium} \label{sec:intro}

The accretion of gas onto freely floating BHs has been analyzed in~\cite{Agol:2001hb} and applied to PBHs in~\cite{Inoue:2017csr}.  Bondi-Hoyle-Lyttleton accretion results in the mass accretion rate\footnote{Recent 3D hydrodynamical simulations show that the accretion rate at high Mach number would be limited to $\sim10$--$20$\% of the canonical Bondi-Hoyle-Lyttleton accretion rate of Eq.~\eqref{eq:bondihoyle}~\cite{Guo:2020qkf}. Since this depends on the assumed conditions and the dominant constraints in our study rely on low-velocity regime, throughout this work we adopt the canonical rate.}~\cite{1939PCPS...35..405H,1944MNRAS.104..273B,1952MNRAS.112..195B}
\begin{align}
\label{eq:bondihoyle}
    \dot{M} = 4 \pi r_B^2 \tilde{v} \rho = \frac{4\pi G^2 M^2 n \mu m_p}{\Tilde{v}^3}~,
\end{align}
where $M$ is the PBH mass, $r_B = G M/\tilde{v}^2$ is the Bondi radius, $\mu$ is the mean molecular weight, $n$ is the ISM gas number density, $m_p$ is the proton mass and
 $\tilde{v} \equiv (v^2+c_s^2)^{1/2}$. Here, $v$ is the PBH speed relative to the gas and $c_s$ is the temperature-dependent sound speed in gas, which we take to be $c_s \sim 10$~km/s~\citep{Inoue:2017csr}.  
 
The accretion rate can be related to the bolometric emission luminosity as $L = \epsilon(\dot{M})\dot{M}$, 
where $\epsilon(\dot{M})$ is the radiative efficiency which scales with accretion rate. The Eddington accretion rate, assuming a characteristic radiative efficiency of $\epsilon_0 = 0.1$, is defined in terms of the Eddington luminosity $\dot{M}_\textrm{Edd} = L_\textrm{Edd}/\epsilon_0 c^2$. A convenient parameter for characterizing the accretion flow is $\dot{m} = \dot{M}/\dot{M}_\textrm{Edd}$.

With a sufficient angular momentum, the  infalling gas can form an accretion disk around the BH. The angular momentum necessary for a disk formation can be supplied by perturbations in the density or the velocity of the accreting gas. For a Schwarzschild BH, the inner radius of the disk is taken to be the innermost stable circular orbit (ISCO) of a test particle  $r_{\rm ISCO} = 3 r_s$, where $r_s = 2 G M/c^2$ is the Schwarzschild radius. Following the arguments of~\cite{Agol:2001hb,Inoue:2017csr}, we have confirmed that an accretion disk always forms for our parameters of interest.

If PBHs constitute a fraction $f_{\rm PBH}$ of the DM, the total number of PBHs of mass $M$ within a volume $V$  is 
\begin{equation}
N_{\rm PBH}(M) = f_{\rm PBH} \frac{\rho_{\rm DM}V}{M}~, 
\end{equation}
where $\rho_{\rm DM}$ is the DM density, assumed to be approximately constant. 
We assume a monochromatic PBH mass function for definiteness and for presenting our results in the form of a differential exclusion plot. The velocity of PBHs contributing to the DM can be described by a Maxwell-Boltzmann distribution 
\begin{equation}
\label{eq:maxwellian}
    f_v(v) = \sqrt{\frac{2}{\pi}}\frac{v^2}{\sigma_v^3}\exp\left(-\frac{v^2}{2\sigma_v^2}\right)~,
\end{equation}
where $\sigma_v$ is the velocity dispersion in a given system. A distribution in gas number density $f_n(n)$ can also be introduced, as in~\cite{Agol:2001hb, Inoue:2017csr}.

\section{Gas Heating} \label{sec:intro}

For a gas system in thermal equilibrium, the total amount of heating by PBHs of mass $M$ is
\begin{align}
\label{eq:totalheateq}
    H_{\rm tot}(M) =&~ N_{\rm PBH}(M) H(M) \\
    =&~ \int_{n_\textrm{min}}^{n_\textrm{max}}\int_{v_\textrm{min}}^{v_\textrm{max}} dn dv \frac{df_n}{dn}\frac{df_v}{dv}\mathcal{H}(M,n,v)~, \notag
\end{align}
where $df_n/dn$ is the gas density distribution, $df_v/dv$ is the PBH relative speed distribution and $\mathcal{H}(M,n,v)$ is the amount of heat deposited into the system from a single PBH. Here, $\mathcal{H}$ represents the cumulative contribution from all heating processes. For photon emission and outflows we perform an additional integration to treat the absorption efficiency.
For gas of approximately constant density, one can replace $df_n/dn$ by a  delta function.

First, we consider gas heating due to photon emission from accretion. Emission in the X-ray band generally constitutes the dominant contribution and it becomes more efficient at high mass accretion rates.

Photon emission from accretion depends on the accretion flow. 
To characterize the accretion flow, we follow the scheme outlined in ~\cite{Yuan:2014gma} and assume that the accretion flow results in a (geometrically) thin disk for $\dot{m}>\dot{M}=0.07\alpha$. The thin $\alpha$-disk is the so-called standard disk~\citep{Shakura:1972te}, where $\alpha \sim 0.1$ is a phenomenological parameter describing viscosity. A thin disk is optically thick and efficiently emits blackbody radiation. Thin disk emission allows for a fully analytic description, and we employ the scaling characterization of  ~\cite{Pringle:1981ds}.

For accretion rates with efficiency below the thin disk regime, accretion is described by the advection dominated accretion flow (ADAF)~\citep[][]{Narayan:1994xi,Yuan:2014gma}. Here, the heat generated by viscosity during accretion is not efficiently radiated out, and much of the energy is advected via matter heat capture into the BH event horizon along with the gas inflow. In contrast with the thin disk, the ADAF ``disk'' is geometrically thick and optically thin. 

An ADAF disk results in a complicated multi-component emission spectrum. We consider three components of the ADAF spectrum, arising from electron cooling: synchrotron radiation, inverse Compton scattering and bremsstrahlung. To describe the ADAF spectrum, we employ approximate analytic expressions obtained in ~\cite{Mahadevan:1996jf} in combination with the updated values for the phenomenological input parameters consistent with recent numerical simulations and observations~\citep{Yuan:2014gma}.  We take the ratio of direct viscous heating to electrons and ions $\delta = 0.3$, and the ratio of gas pressure to total pressure $\beta = 10/11$.

We do not consider a slim disk or other solutions for near- or super-Eddington accretion, $\dot{m}\sim 1$, because such high accretion rates are not achieved for PBH masses and gas densities that we discuss. 

Emitted photons heat the ISM. Hydrogen gas is optically thin to radiation below the ionization threshold of $E_i = 13.6~\textrm{eV}$, and the velocity dispersion is not high enough for a significant Doppler broadening of the emission spectra. Thus, we ignore the absorption of photons with energies less than $E_i$. 

If the medium is optically thick, the photons are absorbed, and most of their  energy is deposited as heat. For absorption of photons with $E>E_i$, we use the photo-ionization cross-section~\cite{Bethe1957,1990A&A...237..267B}, $  \sigma(E) = \sigma_0 y^{-\frac{3}{2}}\left(1+y^{\frac{1}{2}}\right)^{-4}$, 
where $y=E/E_0$, $E_0 = 1/2 E_i$ and $\sigma_0 = 
6.06\times10^{-16}\textrm{ cm}^2$. 
The optical depth of a gas system of size $l$ and density $n$ is $\tau (n, E) = \sigma(E) n l$. Above $30\textrm{ eV}$, we use the combined attenuation length data from Fig.~(32.16) of ~\cite{Olive_2014}. 
The resulting heating power is
\begin{equation}
\label{eq:heatingphotons}
    \mathcal{H}_{\rm phot} (M,n,v) = 
    \int_{E_i}^{E_\textrm{max}}L_\nu (M,n,v) f_h \left(1-e^{-\tau}\right)d\nu~,
\end{equation}
where $L_{\nu}(\nu)$ is the luminosity for the corresponding photon emission process and $f_h$ is the fraction of energy deposited as heat that we estimate to be $\sim 1/3$ \cite{2010MNRAS.404.1869F}. For both the ADAF and thin disk regimes, the emission spectrum is exponentially decreasing at high energies, and we evaluate the integral up to the maximum energy $E_{\rm max}=\infty$.

The second contribution to gas heating that we consider is dynamical friction due to gravitational interactions of traversing PBHs with the surrounding medium.  
Dynamical friction can be described as work done by the ``gravitational drag'' force $F_{\rm dyn}$ (see e.g.~\cite{2008gady.book.....B,1999ApJ...513..252O}).  The resulting power deposited as heat is
\begin{equation}
\label{eq:dynpowergen}
    \mathcal{H}_\textrm{dyn} =  F_{\rm dyn} v
    = -\frac{4\pi G^2 M^2 \rho}{v} I~,
\end{equation}
where $G$ is the gravitational constant, $\rho$ is the gas density and $I$ is a velocity-dependent geometrical factor that differs if the medium is collisionless or not~\cite{2008gady.book.....B,1999ApJ...513..252O}. We have confirmed that the effect of the dynamical friction on the PBH velocity is small.

As a third heating component,
mass outflows (winds) composed of protons can also contribute and they are expected to be significant for hot accretion flows~\citep{Yuan:2014gma}. In contrast to jets\footnote{As jets are typically associated with Kerr black holes, they would require a separate treatment and we do not consider them here.}, the outflows are not highly relativistic and cover a wider angular distribution.
The outflows reduce the accretion rate at smaller radii and can be approximately modelled by a self-similar power-law form~\cite{Blandford:1998qn}
\begin{equation}
\label{eq:inflowgen}
    \dot{M}_\textrm{out}(r) = \dot{M}_\textrm{in}(r_\textrm{out})\left(\frac{r}{r_\textrm{out}}\right)^s~,
\end{equation}
where $r_{\rm out}$ is the outer radius and the real index $s$, $0 \leq s < 1$ is limited by energy and mass conservation. There is a significant uncertainty in the description of the outflows. 
We vary the exponent $s$ in the range $0.5-0.7$~\cite{2012ApJ...761..130Y,Yuan:2014gma}, in agreement with numerical simulations. Furthermore, we consider the outer radius $r_\textrm{out}$ over a wide range of values, from $100 r_s$~\cite{Xie:2012rs} to  $r_B$~\cite{2011MNRAS.415.1228P}. The resulting outgoing wind has a velocity that is a fraction $f_k\simeq0.1-0.2$~\cite{2013ApJ...767..105L,2013IAUS..290...86Y,2012ApJ...761..130Y}
 of the Keplerian velocity at the radius at which it is ejected, i.e. $v(r) \simeq f_k \sqrt{G M/r}$.
We note that additional considerations regarding details of accretion may reduce emission efficiency (e.g.~feedback), but we do not expect this to be very significant.

To evaluate how much energy is deposited into the gas system from streaming outflow protons, we convolve the proton emission with the heat generated per proton $\Delta E$. The total heat deposited in the gas system is 
\begin{equation}
\label{eq:outflowint}
    \mathcal{H}_\textrm{out} = 
    \int_{r_\textrm{in}}^{r_\textrm{out}}  \frac{f_h \Delta E}{\mu m_p}\frac{d\dot{M}_\textrm{out}}{dr} dr~,
\end{equation}
where 
\begin{equation}
\label{eq:energydepo}
    \Delta E = \int \frac{dE}{dx} dx \simeq \textrm{min}(E,n S(E) r_\textrm{max})~
\end{equation}
takes into account energy losses due to the  proton stopping power $dE/dx = nS(E)$ adopted from ~\cite{2019PhRvA..99d2701B} (see their Fig.~9). Here, $r_{\max}$ is taken to be the size of the gas system.

\begin{figure*}[tb]
\begin{center}
\includegraphics[trim={5mm 0mm 40 0},clip,width=.475\textwidth]{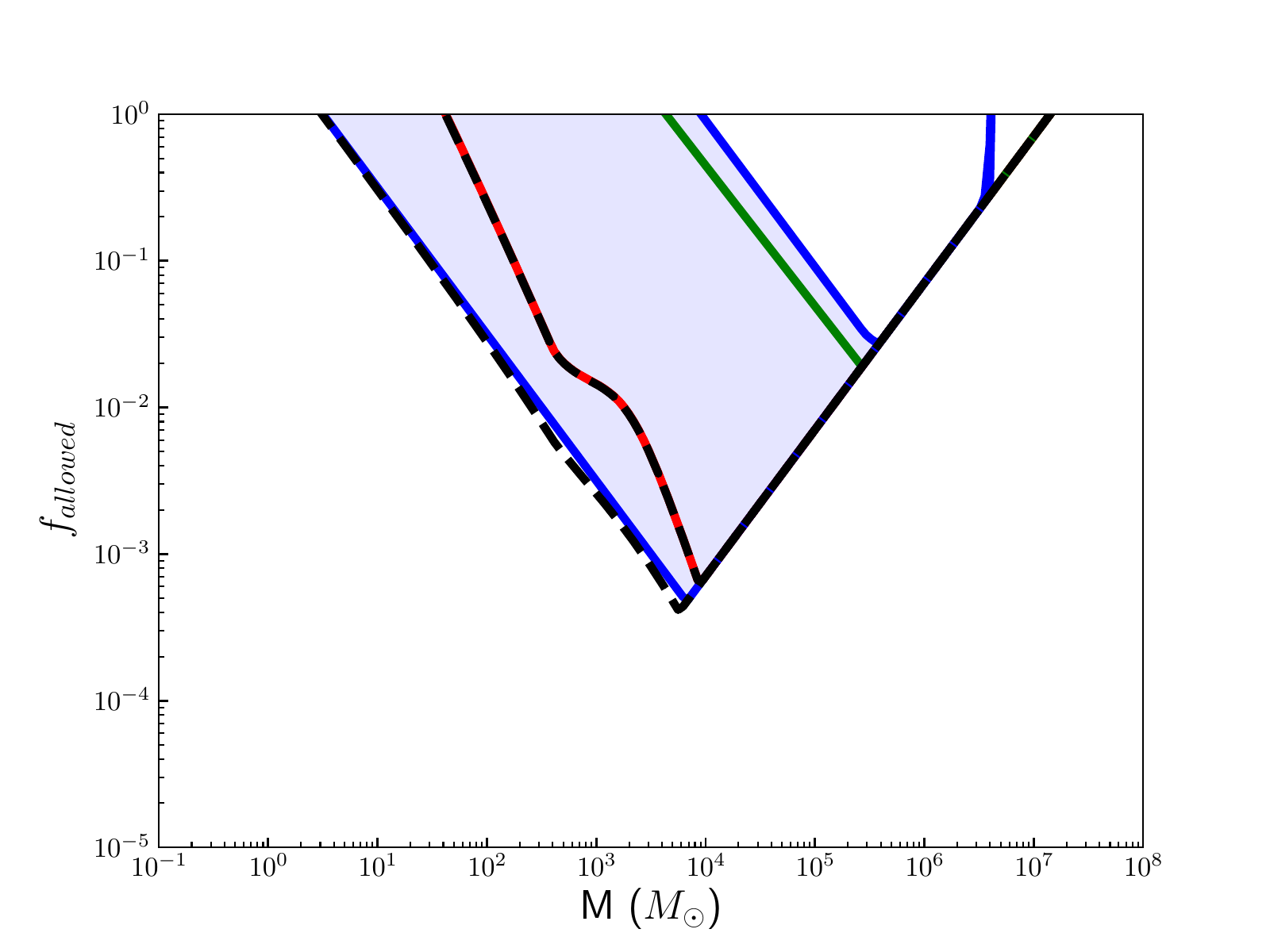}
\includegraphics[trim={5mm 0mm 40 0},clip,width=.475\textwidth]{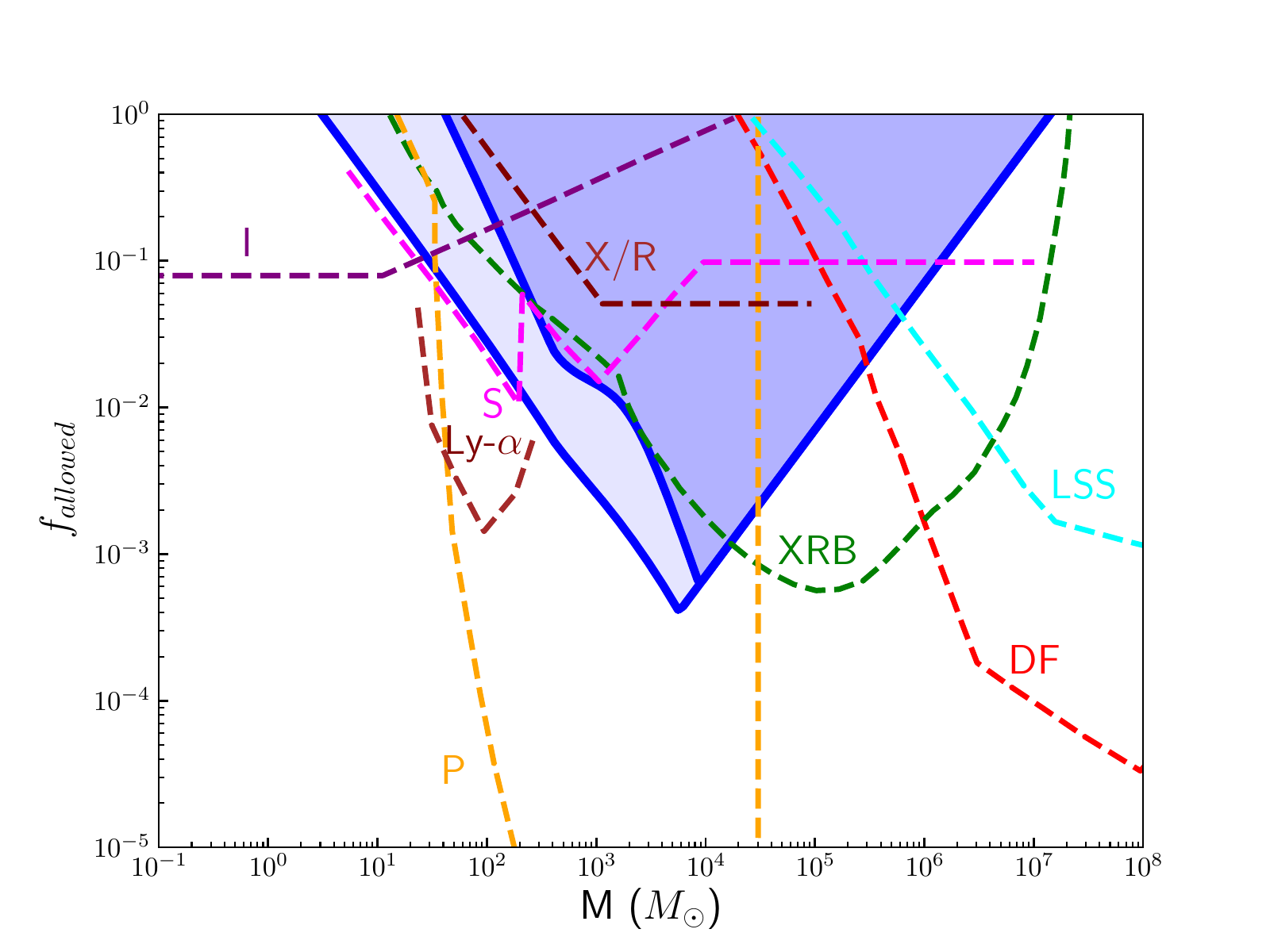}

\caption{\label{fig:leotgraphs}  \textbf{Left:}
Constraints from Leo T on the fraction of DM in PBHs, for a monochromatic mass function, derived from considerations of only photon emission (red), dynamical friction (green), mass outflows (blue), as well as combined heating (dashed black). The reach of the constraints is bounded by the diagonal (solid black) line from the condition of Eq.~\eqref{eq:incredlimit}. The uncertainty in the emission and outflow input parameters leads to the uncertainty in the corresponding constraint (upper and lower dashed black lines). \textbf{Right:} Constraints from the Leo T dwarf galaxy on the PBH gas heating are shown in blue. The light blue shaded band denotes the variation in the PBH emission parameters. Other existing constraints (see Ref.~\citep{Carr:2020gox})  are shown by dashed lines, including Icarus~\cite{2018PhRvD..97b3518O}
(I) caustic crossing in purple, Planck~\cite{Ali-Haimoud:2016mbv,Serpico:2020ehh}
(P) in yellow, X-ray binaries~\cite{Inoue:2017csr}
(XRB) in green, dynamical friction of halo objects
(DF) in red, Lyman-$\alpha$~\cite{2019PhRvL.123g1102M}
(Ly-$\alpha$) in maroon, combined bounds from the survival of astrophysical systems in Eridanus II~\cite{Brandt:2016aco},
 Segue 1~\cite{2017PhRvL.119d1102K},
 and disruption of wide binaries~\cite{2014ApJ...790..159M}
(S) shown in magenta, large scale structure~\cite{Carr:2018rid}
(LSS) in cyan, and X-ray/radio~\cite{Ricotti:2007au}
(X/R) in brown.}
\end{center}
\end{figure*}

\section{Target System: Leo T}

We demonstrate our analysis by applying it to dwarf galaxies, focusing on the Leo T dwarf galaxy.
We stress, however, that our methods are general and can be readily applied to other systems as well. To constrain the PBH mass fraction $f_\textrm{PBH}$, we consider the balance between the heating and cooling processes of the gas system. Our approach to set the limits is similar to that  used for particle  DM~\citep{Bhoonah:2018wmw,Farrar:2019qrv,Wadekar:2019xnf}, but the heating mechanisms and the preferred gas systems are different in our case. 
 
For simplicity, we ignore the contribution of natural heating sources (e.g. stellar radiation), and hence our bounds are conservative. 
Requiring thermal equilibrium, we only consider gas systems that are expected to be approximately stable on sufficiently long timescales $\tau_\textrm{sys}$. Hence, the characteristic time over which the gas system remains steady must be greater than the cooling timescale of the gas $\tau_\textrm{therm}$, i.e.~$\tau_\textrm{sys} \gg \tau_\textrm{therm} = 3nkT/2\dot{C}$,
where $k$ is the Boltzmann constant and $\dot{C}$ is the gas cooling rate per volume. We note that
presence of magnetic fields in the ISM can also affect 
emitted protons that we consider.
 However, the strength, orientation
and distribution of magnetic fields in Leo T is highly uncertain and very poorly known.
Hence, we do not consider such effects.

Gas temperature exchange is a complex process, and a detailed analysis involving a full chemistry network can be performed using numerical  methods~\citep{Smith:2016hsc}. For the  parameters of interest, we employ  approximate results obtained in ~\cite{Wadekar:2019xnf}. For hydrogen gas, the cooling rate is given by
\begin{equation}
\label{eq:coolinggen}
    \dot{C} = n^2 10^{[\textrm{Fe/H}]}\Lambda(T)~,
\end{equation}
where [Fe/H]$ \equiv \log_{10}(n_{\rm Fe}/n_{\rm H})_{\rm gas} - \log_{10}(n_{\rm Fe}/n_{\rm H})_{\rm Sun}$ is the metallicity, and $\Lambda(T) \propto 10^{[{\rm Fe/H}]}$ is the cooling function. Fitting numerically to the results of ~\cite{Smith:2016hsc} library, one can obtain $\Lambda(T) = 2.51\times10^{-28}T^{0.6}$, valid for $300~\text{K} < T < 8000~\text{K}$~\citep{Wadekar:2019xnf}.

The total PBH heating in the cloud of gas $H_{\rm tot} = N_\textrm{PBH} H(M)= f_\textrm{PBH}\rho_\textrm{DM}V H(M)/M$ given by Eq.~\eqref{eq:totalheateq}, where $H(M)$ is the average heat generated from one PBH of mass $M$, should be less than the total cooling $\dot{C} V$. This yields a condition on the PBH abundance that we use to set our limits: 
\begin{equation}
\label{eq:genbound}
    f_\textrm{PBH} < f_\textrm{bound} =\frac{M\dot{C}}{\rho_\textrm{DM}H(M)}~.
\end{equation}

We note that gas heating can be used to  set a limit on the PBH abundance only if it is statistically likely for the gas system to harbor PBHs. If the PBH number density is so low that, on average, a gas system of the size $r_{\rm sys}$ contains fewer than one PBH, i.e., $f_{\rm PBH}\, \rho_{\rm DM} (4\pi r_{\rm sys}^3/3)/M<1$, such a system cannot be used for our purposes.  We, therefore, set a limit only as long as 
\begin{equation}
\label{eq:incredlimit}
    f_\textrm{bound} > \frac{3M}{4\pi r_\textrm{sys}^3 \rho_\textrm{DM}}.
\end{equation}

The gas in the inner region of Leo T, at a radius $r\lesssim 350 \textrm{ pc}$ from its center, is dominated by atomic hydrogen, while the gas outside is highly ionized~\citep{2013ApJ...777..119F}. Since the free electrons in the ionized region cool very efficiently~\citep{Smith:2016hsc}, we limit our analysis to the central region of Leo T. From the model of ~\cite{2013ApJ...777..119F}, the hydrogen gas density is found to vary from $\sim0.2 \textrm{ cm}^{-3}$ in the center to $\sim0.03 \textrm{ cm}^{-3}$ at $r=350\textrm{ pc}$. Both the cooling and heating rates scale roughly as $n^2$, so we approximate the gas density to be a constant $n=0.07\textrm{ cm}^{-3}$ in the inner region.
Similarly, we approximate the DM mass density to be a constant value of $1.75 \textrm{ GeV}/\textrm{cm}^{3}$. The hydrogen gas has a dominant non-rotating warm component with a velocity dispersion of $\sigma_g=6.9~ \textrm{km}/\textrm{s}$ and $T\simeq 6000$~K~\citep{2008MNRAS.384..535R,2013ApJ...777..119F} and also a sub-dominant cold component that we ignore. The DM is expected to have the same velocity dispersion as the gas, $\sigma_v=\sigma_g$. The sound speed is taken to be $c_s=9\textrm{ km}/\textrm{s}$ from the adiabatic formula with $T\simeq 6000$~K. Combining the radius and number density, the column density of hydrogen gas in the central region of Leo T is $n r_\textrm{sys} = 7.56\times10^{19}\textrm{ cm}^{-2}$. We adopt the gas metallicity to approximately follow the stellar one\footnote{This is accurate to factor of few.}, $\textrm{[Fe/H]} \simeq -2$ estimated by stellar spectra~\citep{2008ApJ...685L..43K}. Using the above parameters in Eq.~\eqref{eq:coolinggen}, the resulting Leo T's cooling rate is taken to be $\dot{C} = 2.28\times10^{-30}\textrm{ erg}\textrm{ cm}^{-3}\textrm{ s}^{-1}$.

In Fig.~\ref{fig:leotgraphs} we display the resulting limits from gas heating in Leo T on PBHs contributing to DM, along with other existing constraints.

\section{Summary}

In summary, we have presented a new constraint on the abundance of PBHs over a broad  mass range of $\mathcal{O}(1) M_{\odot}-10^7 M_{\odot}$. This parameter space covers the detected stellar and the very recently observed intermediate-mass BHs, as well as seeds for supermassive black holes. PBH interactions with ISM result in the heating of gas, which we applied to dwarf galaxy Leo T to set the limit.  We considered several generic heating mechanisms, including the photon emission from accretion, dynamical friction, and mass outflows/winds. This type of a constraint has not been previously considered for PBHs. Our limit does not depend on the cosmological history, which makes it an attractive independent test of PBHs in the IMBH mass-range. Our analysis can be readily applied to other systems.

\acknowledgments

The work of G.B.G., A.K., V.T., and P.L. was supported in part by the U.S. Department of Energy (DOE) grant No. DE-SC0009937. This work was supported in part by the MEXT
Grant-in-Aid for Scientific Research on Innovative Areas (No. 20H01895 for K.H.). A.K. was also supported by Japan Society for the Promotion of Science (JSPS) KAKENHI grant No.
JP20H05853. Y.I. is supported by JSPS KAKENHI grant No. JP18H05458, JP19K14772, program of Leading Initiative for Excellent Young Researchers, MEXT, Japan, and RIKEN iTHEMS Program. A.K., V.T., and Y.I. are also supported by the World Premier International Research Center Initiative (WPI), MEXT, Japan.

\bibliography{bibliography}{}

\end{document}